\documentclass[reprint, superscriptaddress, preprintnumbers, aps, prl]{revtex4-1}

\usepackage{setup}% all includes & macros
\DeclareRobustCommand{\ptv}{{\ensuremath{p_\rT^V}}\xspace}
\DeclareRobustCommand{\ptw}{{\ensuremath{p_\rT^\PW}}\xspace}
\DeclareRobustCommand{\ptz}{{\ensuremath{p_\rT^\PZ}}\xspace}
\DeclareRobustCommand{\ptcut}{{\ensuremath{p_{\rT,\cut}^V}}\xspace}

\begin{document}

\preprint{CERN-TH-2017-280, IPPP/17/109, ZU-TH 38/17}

\title{NNLO QCD corrections to the transverse momentum \\ distribution of weak gauge bosons}%

\author{A.\ Gehrmann--De Ridder}
 \affiliation{Institute for Theoretical Physics, ETH, CH-8093 Zürich, Switzerland}%
 \affiliation{Department of Physics, University of Zürich, CH-8057 Zürich, Switzerland}%
\author{T.\ Gehrmann}%
 \affiliation{Department of Physics, University of Zürich, CH-8057 Zürich, Switzerland}%
\author{E.\ W.\ N.\ Glover}%
 \affiliation{Institute for Particle Physics Phenomenology, Durham University,  Durham DH1 3LE, UK}%
\author{A.\ Huss}%
 \affiliation{Theoretical Physics Department, CERN, Geneva, Switzerland}%
\author{D.\ M.\ Walker}%
 \affiliation{Institute for Particle Physics Phenomenology, Durham University,  Durham DH1 3LE, UK}%

\date{\today}% It is always \today, today,

\begin{abstract}
The transverse momentum spectra of weak gauge bosons and their ratios probe the underlying dynamics and are crucial in testing our understanding of the Standard Model.
They are an essential ingredient in precision measurements, such as the $\PW$-boson mass extraction.
To fully exploit the potential of the LHC data, we compute the second-order (NNLO) QCD corrections to the inclusive-$\ptw$ spectrum as well as to the ratios of spectra for $\PWm/\PWp$ and $\PZ/\PW$.
We find that the inclusion of NNLO QCD corrections considerably improves the theoretical description of the experimental CMS data and results in a substantial reduction of the residual scale uncertainties.  
\end{abstract}

\maketitle

%--------------------------------------------------------------------
\section{\label{sec:intro}Introduction}
%--------------------------------------------------------------------

%%-- motivation: general DY 
The production of electroweak (EW) gauge bosons with subsequent leptonic decay, known as the Drell--Yan process, is one of the most prominent processes at hadron--hadron colliders such as the LHC.
Not only are the gauge bosons produced in abundance, but the clean leptonic signature allows this class of processes to be measured with great precision. 
As a consequence, the Drell--Yan-like production of $\PW$ and $\PZ$ bosons is among the most important ``standard candles'' at hadron colliders and, as such, has a wide range of applications. 

%%-- motivation: pt(V) 
The transverse-momentum spectrum of the gauge bosons ($\ptv$) takes a particularly important rôle in this respect:
Different kinematical regimes of this observable probe various aspects of the predictions, such as resummation and non-perturbative effects at low $\ptv$, fixed-order predictions at intermediate to high $\ptv$, and also electroweak Sudakov logarithms at very high $\ptv$.
As such, detailed theory--data comparisons of this observable constitute crucial probes to test the Standard Model predictions.
The $\ptv$ distribution can also provide important constraints in the fit of parton distribution functions (PDFs), as was studied in Ref.~\cite{Boughezal:2017nla} for the case of the $\PZ$-boson spectrum.

%%-- motivation: ratios 
Ratios between two $\ptv$ spectra of different processes, such as $\PWm/\PWp$ and $\PZ/\PW$, shed further light on the composition of the proton and are also important inputs to precision measurements.
Most notably, a precise theoretical understanding of the ratio between $\PZ$- and $\PW$-boson production cross sections is of critical importance in the measurement of the $\PW$-boson mass~\cite{Aaltonen:2013iut,Aaboud:2017svj}, where the modeling of the $\PW$-boson $\ptv$ spectrum is obtained indirectly through $\ptz$.

%%-- experimental overview: pt(V) & ratio
The $\ptv$ spectra of weak gauge bosons have been studied by the CDF~\cite{Aaltonen:2012fi} and D0~\cite{Abbott:1998jy,Abbott:1999wk,Abbott:1999yd,Abbott:2000xv,Abazov:2001nta,Abazov:2007ac,Abazov:2010kn} collaborations at the Tevatron collider. 
Corresponding measurements at the LHC have been performed by the ATLAS~\cite{Aad:2011fp,Aad:2015auj}, CMS~\cite{Chatrchyan:2011wt,Khachatryan:2015oaa,Khachatryan:2016nbe}, and LHCb~\cite{Aaij:2015gna,Aaij:2016mgv} experiments and constitute an integral part of the current and future LHC programme.
CMS also studied the ratios of the $\ptv$ spectra for the $\PWm/\PWp$ and $\PZ/\PW$ processes~\cite{Khachatryan:2016nbe}.

%%-- theory overview: just fixed-order to V+jet
In view of the experimental precision that is achievable for the measurement of the $\ptv$ spectra, it becomes mandatory to have theoretical predictions of the highest possible accuracy.
Given that a non-vanishing value for the transverse momentum implicitly requires a balancing recoil, the $\ptv$ spectrum above a finite value, $\ptv>\ptcut$, is closely related to $V+\jet$ production. 
In the context of fixed-order calculations, predictions are known to next-to-leading order (NLO) accuracy for this process class in perturbative QCD~\cite{Giele:1993dj} and electroweak effects~\cite{Kuhn:2005az,*Kuhn:2007qc,*Denner:2009gj,*Denner:2011vu}. 
In recent years, substantial progress has been made in the computation of predictions at one order higher in pQCD and the next-to-next-to-leading order (NNLO) QCD corrections are now available for $\PW+\jet$~\cite{Boughezal:2015dva}, $\PZ+\jet$~\cite{Ridder:2015dxa,Boughezal:2015ded}, and $\Pgg+\jet$~\cite{Campbell:2016lzl,*Campbell:2017dqk} production.
In this work, we present the calculation of the $\order{\alphas^3}$ NNLO QCD corrections to $\PW$ production at finite transverse momentum with leptonic decay, 
\begin{equation}
  \label{eq:procW}
  \Pp+\Pp \to \PWpm (\to \Pl + \Pgnl)\bigr\rvert_{\ptv>\ptcut} + X ,
\end{equation}
which is closely related to the $\PZ$ transverse momentum distribution discussed in Ref.~\cite{Ridder:2016nkl,*Gehrmann-DeRidder:2016jns,*Gauld:2017tww}.
These corrections receive contributions from three classes of parton-level processes with different partonic multiplicities:
(a) the two-loop $\PW$-boson-plus-three-parton processes~\cite{Garland:2001tf,*Garland:2002ak,*Gehrmann:2011ab}, (b) the one-loop $\PW$-boson-plus-four-parton processes~\cite{Glover:1996eh,*Bern:1996ka,*Campbell:1997tv,*Bern:1997sc}, and (c) the tree-level $\PW$-boson-plus-five-parton processes~\cite{Hagiwara:1988pp,*Berends:1988yn}. 
All three types of contributions are infrared divergent and only their sum is finite. 
To this end, we employ the antenna subtraction formalism~\cite{GehrmannDeRidder:2005cm,*GehrmannDeRidder:2005aw,*GehrmannDeRidder:2005hi,*Daleo:2006xa,*Daleo:2009yj,*Boughezal:2010mc,*Gehrmann:2011wi,*GehrmannDeRidder:2012ja,*Currie:2013vh} for the cancellation of infrared divergences.
We further provide predictions for ratios between different weak boson processes.

%--------------------------------------------------------------------
\section{\label{sec:calc}Details of the Calculation}
%--------------------------------------------------------------------

%%-- CMS setup
We adopt the setup of the CMS measurement of Ref.~\cite{Khachatryan:2016nbe} and perform a comparison of the predictions for the normalized $\ptv$ distributions for $\PW$- and $\PZ$-boson production and their ratios.
The measurement is performed in the fiducial volume defined by the lepton cuts $p_\rT^\Pe>25~\GeV$ ($p_\rT^\Pgm>20~\GeV$) and $\lvert\eta^\Pe\rvert<2.5$ ($\lvert\eta^\Pgm\rvert<2.1$) for the electron (muon) channel.
For the neutral-current process, an additional invariant-mass cut of $60<m_{\Pl\Pal}<120~\GeV$ is imposed on the lepton-pairs. 

%%-- NNLOJET setup
The transverse-momentum distributions shown here are $\order{\alphas^3}$, where final-state QCD emissions are treated fully inclusively while imposing a transverse-momentum cut of $\ptv>7.5~\GeV$ on the vector bosons.
This cut renders the calculation infrared finite as it enforces the presence of final-state partons that recoil against the vector boson.
It is further aligned with the upper edge of the first $\ptv$ bin of the charged-current and ratio measurements.
The normalization in the normalized distributions is obtained from the inclusive Drell--Yan calculation, which we evaluate at $\order{\alphas^2}$ throughout. 
All these processes are implemented in the flexible parton-level Monte Carlo generator \textsc{NNLOjet}.
It provides the necessary infrastructure for the antenna subtraction formalism, used to redistribute and cancel the infrared divergences appearing in contributions of different parton multiplicities.
This program combines all parton-level sub-processes contributing at a given order in $\alphas$ and further allows to provide fully differential results in the form of binned distributions which can directly be compared to LHC data.
For the PDFs, we employ the central member of the \verb|NNPDF31_nnlo|~\cite{Ball:2017nwa} set with $\alphas(M_\PZ)=0.118$ for all predictions at LO, NLO, and NNLO.

%%-- scale setting and variation
In order to assess the theory uncertainties, we independently vary the factorization ($\muf$) and renormalization ($\mur$) scales by factors of $\tfrac{1}{2}$ and $2$ around the central scale $\mu_0$, while imposing the restriction $\tfrac{1}{2}\leq \muf/\mur \leq 2$.
The central scale choice is given by the transverse energy
\begin{equation}
  \label{eq:mu0}
  \mu_0 = E_\rT \equiv \sqrt{M_V^2 + (\ptv)^2} ,
\end{equation}
where $M_V$ and $\ptv$ denote the invariant mass and transverse momentum of the final-state lepton pair.
For the ratios and double-ratios encountered in the normalized distributions and their ratios, we generalize this procedure and consider the uncorrelated variation of all scales appearing inside the different parts while imposing $\tfrac{1}{2}\leq \mu/\mu^{\prime} \leq 2$ between all pairs of scales.
This prescription results in a total of 31 and 691 points in the scale variation of the normalized distributions and their ratios, respectively.

%--------------------------------------------------------------------
\section{\label{sec:res}Results}
%--------------------------------------------------------------------

%%-- normalized distributions
\begin{figure}[t]
  \includegraphics{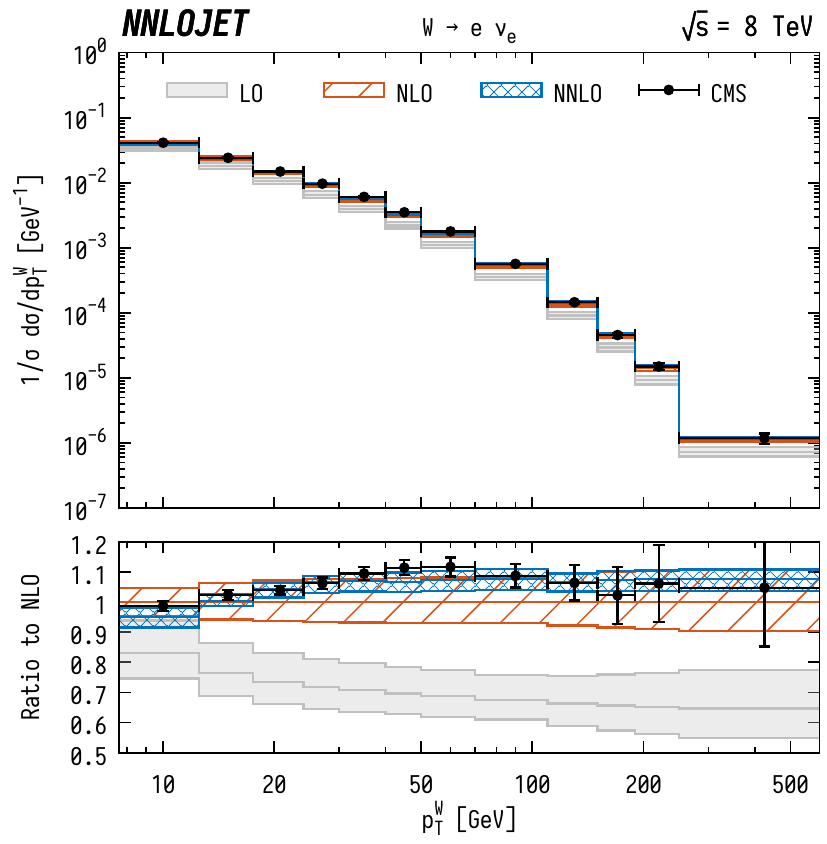}
  \caption{\label{fig:W_el_ptnorm} 
    Normalized $\ptw$ distribution for $\PW=\PWp+\PWm$ production with subsequent decay into electrons.
    Predictions at LO (gray fill), NLO (orange hatched), and NNLO (blue cross-hatched) are compared to CMS data from Ref.~\cite{Khachatryan:2016nbe}.
    The bands correspond to scale uncertainties estimated as described in the main text.
  }
\end{figure}
\begin{figure}[t]
  \includegraphics{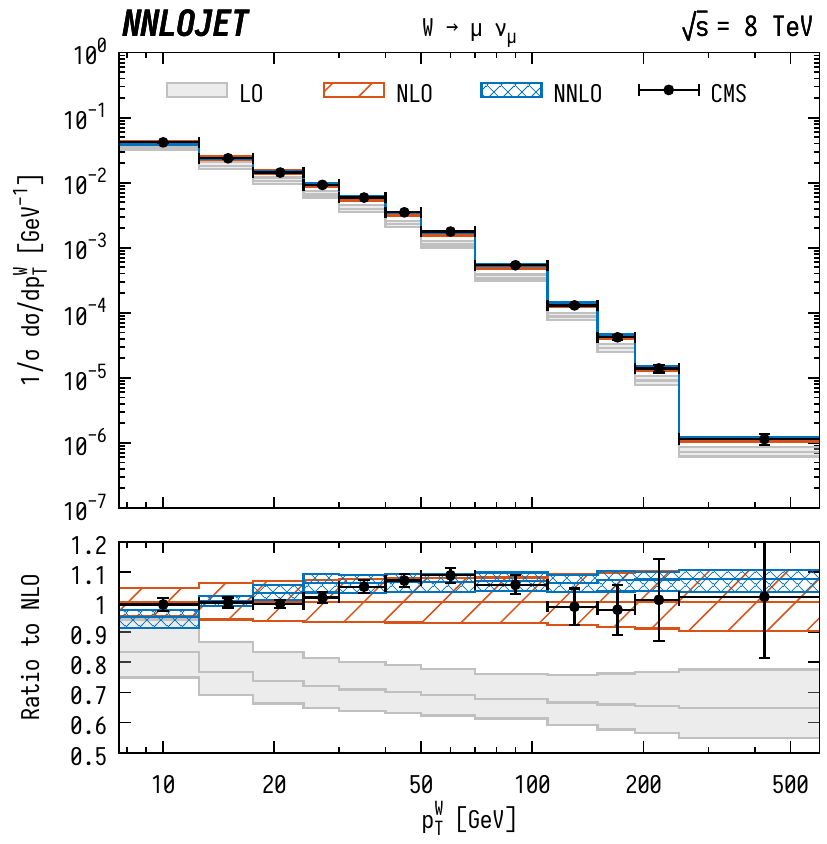}
  \caption{\label{fig:W_mu_ptnorm} 
    Normalized $\ptw$ distribution for $\PW=\PWp+\PWm$ production with subsequent decay into muons.
    Predictions at LO (gray fill), NLO (orange hatched), and NNLO (blue cross-hatched) are compared to CMS data from Ref.~\cite{Khachatryan:2016nbe}.
    The bands correspond to scale uncertainties estimated as described in the main text.
  }
\end{figure}
\begin{figure}[t]
  \includegraphics{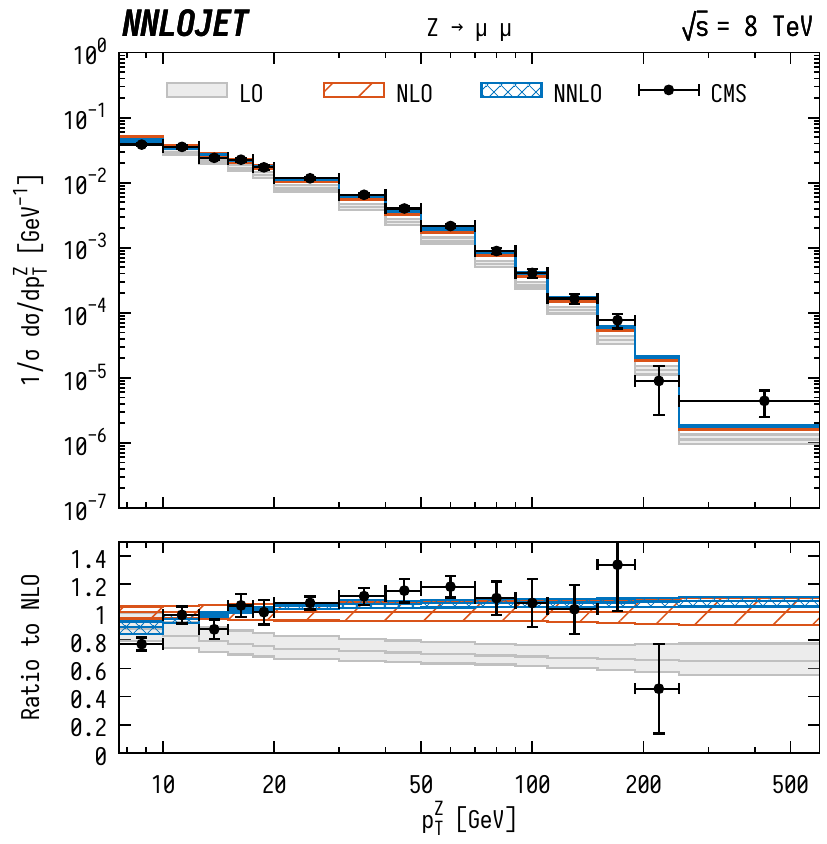}
  \caption{\label{fig:Z_mu_ptnorm} 
    Normalized $\ptz$ distribution for $\PZ$ production with subsequent decay into muons.
    Predictions at LO (gray fill), NLO (orange hatched), and NNLO (blue cross-hatched) are compared to CMS data from Ref.~\cite{Khachatryan:2016nbe}.
    The bands correspond to scale uncertainties estimated as described in the main text.
  }
\end{figure}
Figures~\ref{fig:W_el_ptnorm} and \ref{fig:W_mu_ptnorm} show the normalized transverse-momentum distribution of the $\PW$ boson in the electron and muon channels, respectively. 
In the following, the label ``$\PW\to\Pl\Pgn_\Pl$'' denotes the sum of both the $\PWm\to\Plm\Pagn_\Pl$ and $\PWp\to\Plp\Pgn_\Pl$ processes. 
The NLO corrections are between $10$--$40\%$ with residual scale uncertainties at the level of around $\pm10\%$.
Although the scale-uncertainty bands at NLO mostly cover the experimental data points, systematic differences in the shape between data and the central theory prediction are visible.
In view of the experimental precision, this clearly demonstrates the necessity of higher-order predictions with smaller uncertainties in order to discriminate such behaviors.
The NNLO corrections are positive and between $5$--$10\%$ in the intermediate- to high-$\ptw$ region.
Towards lower $\ptw$, the NNLO corrections become smaller and turn negative in the lowest-$\ptw$ bin. 
The residual scale uncertainties reduce to the level of about $\pm2\%$ and overlap with the NLO scale bands, exhibiting good perturbative convergence.
Most notably, we observe that the shape distortion induced by the NNLO corrections brings the central predictions in line with the measured distributions.

The corresponding comparison for the $\PZ$-boson spectrum is shown in Fig.~\ref{fig:Z_mu_ptnorm}, where the measurement was only performed in the muon channel.
As in the charged-current case, there is a substantial reduction in the scale uncertainties accompanied with an improved description of the shape.
We note that the $y$-range of the bottom panel in Fig.~\ref{fig:Z_mu_ptnorm} has been increased compared to the respective figures of the charged-current process in order to accommodate the experimental data which exhibit larger statistical fluctuations.

%%-- ratios
\begin{figure}
  \includegraphics{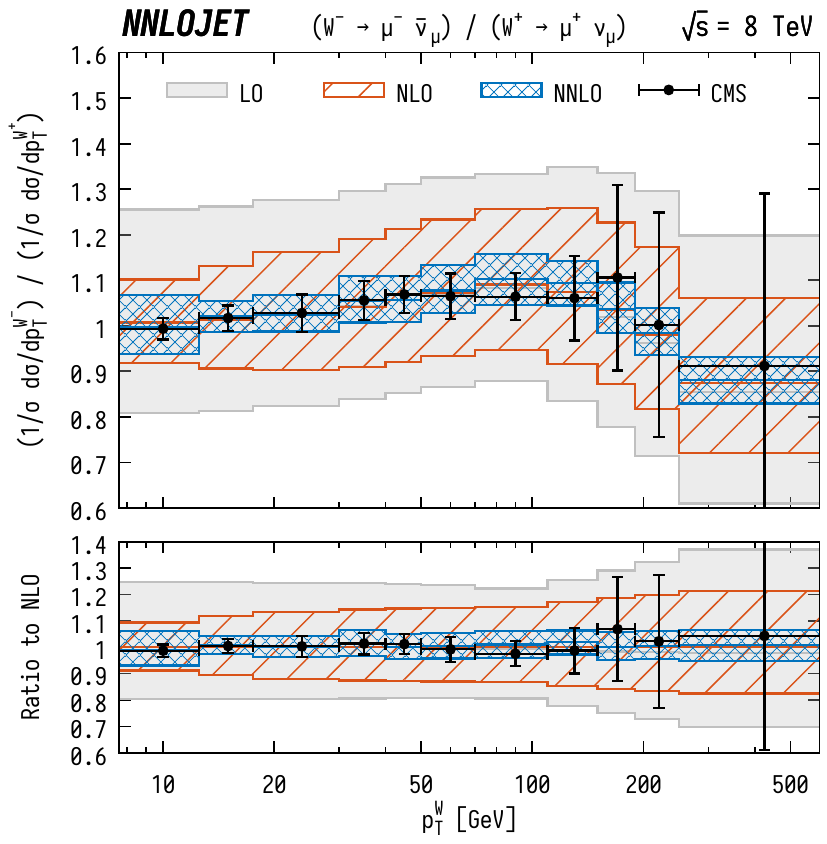}
  \caption{\label{fig:Wm_Wp_ratio} 
    Ratio of normalized $\ptw$ distributions between $\PWm$ and $\PWp$ production in the muon channel.
    Predictions at LO (gray fill), NLO (orange hatched), and NNLO (blue cross-hatched) are compared to CMS data from Ref.~\cite{Khachatryan:2016nbe}.
    The bands correspond to scale uncertainties estimated as described in the main text.
  }
\end{figure}
\begin{figure}
  \includegraphics{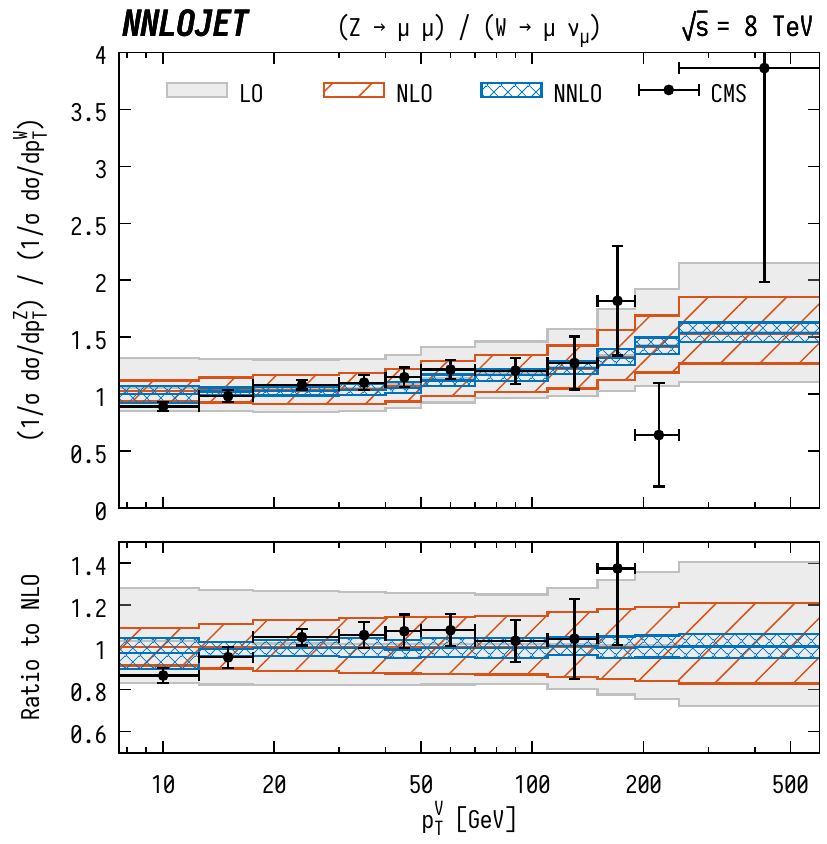}
  \caption{\label{fig:Z_W_ratio} 
    Ratio of normalized $\ptv$ distributions between $\PZ$ and $\PW=\PWp+\PWm$ production in the muon channel.
    Predictions at LO (gray fill), NLO (orange hatched), and NNLO (blue cross-hatched) are compared to CMS data from Ref.~\cite{Khachatryan:2016nbe}.
    The bands correspond to scale uncertainties estimated as described in the main text.
  }
\end{figure}
Figure~\ref{fig:Wm_Wp_ratio} shows the ratio between the normalized distributions of the $\PWm\to\Plm\Pagn_\Pl$ and $\PWp\to\Plp\Pgn_\Pl$ processes.
The ratio is close to one in the lowest $\ptw$ bin and rises up to $\sim1.1$ at $\ptw\approx150~\GeV$, where it turns over and slowly decreases to $0.9$ at $\ptw=500~\GeV$.
The central predictions remain remarkably stable between the different orders, resulting in $K$-factors that are very close to one.
However, the precision of the theory prediction is substantially improved by going to higher orders:
While the scale uncertainties at NLO are between $\pm10$--$20\%$, the NNLO corrections reduce the uncertainties to the level of $\pm5\%$ across most of the $\ptw$ range, never exceeding $\pm10\%$.

The ratio between the $\PZ$- and $\PW$-boson processes are shown in Fig.~\ref{fig:Z_W_ratio}. 
Here, the ratio is again close to one in the low-$\ptv$ bin and shows a steady increase towards higher $\ptv$, reaching about $1.5$ at $\ptv\sim500~\GeV$.
As was the case for the $\PWm/\PWp$ ratio, the QCD corrections are very stable and leave the central predictions largely unaffected, supporting the expected similarity of QCD corrections between $\PZ$ and $\PW$ production.
The higher-order corrections however have a big impact on the scale uncertainties, which are reduced by more than a factor of two across almost all $\ptv$-bins by going from NLO to NNLO and are at the level of $\pm5$--$10\%$.

%--------------------------------------------------------------------
\section{\label{sec:concl}Conclusions}
%--------------------------------------------------------------------

In summary, we have presented the genuine NNLO $\order{\alphas^3}$ corrections to $\ptv$ spectra in inclusive gauge-boson production and their ratios. 
For the first time, corrections at this order have been computed for the inclusive-$\ptw$ spectrum and for ratios of different $\ptv$ spectra.
The latter comprise ratios between the processes $\PWm/\PWp$ and $\PZ/\PW$, which are key ingredients for precision studies such as the $\PW$-boson mass measurement.

We find that the corrections to the transverse momentum distributions of the weak gauge bosons are quite sizable and at the level of $5$--$10\%$.
The inclusion of the NNLO corrections substantially reduces scale uncertainties and further induces changes to the shape that result in a better theory--data agreement. 
Higher-order QCD corrections to the $\PW$- and $\PZ$-boson production processes exhibit a similar behaviour, as it is further supported by the observed corrections to the ratios. 
Here, we find remarkable stability of the central prediction between the different perturbative orders with $K$-factors very close to one.
However, the inclusion of the NNLO corrections are crucial in reducing the theory uncertainties as estimated through the variation of factorization and renormalization scales. 
The observed reduction in scale uncertainties from NLO to NNLO is typically larger than a factor of two, with a residual uncertainty of about $5\%$.

The calculation presented in this work paves the way towards stress-testing Standard Model predictions using the precise experimental data that are available for the $\ptv$ spectra and related observables and to reduce theory uncertainties in the extraction of PDFs and parameters such as $M_\PW$.

\begin{acknowledgments}
The authors thank Xuan Chen, Juan Cruz-Martinez, James Currie, Rhorry Gauld, Marius Höfer, Imre Majer, Tom Morgan, Jan Niehues, and Joao Pires for useful discussions and their many contributions to the \textsc{NNLOjet} code.
This research was supported in part by the UK Science and Technology Facilities Council, by the Swiss National Science Foundation (SNF) under contracts 200020-175595, 200021-172478, and CRSII2-160814, by the Research Executive Agency (REA) of the European Union under the Grant Agreement PITN-GA-2012-316704 (``HiggsTools'') and the ERC Advanced Grant MC@NNLO (340983).
\end{acknowledgments}

%%-- Produces the bibliography via BibTeX.
\bibliography{Vjets_pt}

\end{document}